# Synthesis of SnTe Nanoplates with {100} and {111} Surfaces

*Jie Shen[1], Yeonwoong Jung[1], Ankit S. Disa[2], Fred J. Walker[2], Charles H. Ahn[1,2], Judy J. Cha[1,3]*

[1] Department of Mechanical Engineering and Materials Science, Yale University, New Haven, CT, 06511

[2] Department of Applied Physics, Yale University, New Haven, CT, 06511

[3] Energy Sciences Institute, Yale University West Campus, West Haven, CT, 06477

ABSTRACT

SnTe is a topological crystalline insulator that possesses spin-polarized, Dirac-dispersive surface states protected by crystal symmetry. Multiple surface states exist on the {100}, {110}, and {111} surfaces of SnTe, with the band structure of surface states depending on the mirror symmetry of a particular surface. Thus, to access surface states selectively, it is critical to control the morphology of SnTe such that only desired crystallographic surfaces are present. Here, we grow SnTe nanostructures using vapor-liquid-solid and vapor-solid growth mechanisms. Previously, SnTe nanowires and nanocrystals have been grown.[1-4] In this report, we demonstrate synthesis of SnTe nanoplates with lateral dimensions spanning tens of microns and thicknesses of a hundred nanometers. The top and bottom surfaces are either (100) or (111), maximizing topological surface states on these surfaces. Magnetotransport on these SnTe nanoplates shows high bulk carrier density, consistent with bulk SnTe crystals arising due to defects such as Sn vacancies. In addition, we observe a structural phase transition in these nanoplates from the high temperature rock salt to low temperature rhombohedral structure. For nanoplates with very high carrier density, we observe a slight upturn in resistance at low temperatures, indicating electron-electron interactions.

KEYWORDS. Topological Crystalline Insulator, SnTe, Nanoplates, Vapor-Liquid-Solid Growth

MAIN TEXT

Since the discovery of CdTe/HgTe quantum wells as a two-dimensional (2D) topological insulator (TI) system,[5, 6] prediction and verification of materials as TIs has been expanding.[7, 8] Most widely studied among them are the layered chalcogenides, $Bi_2Se_3$ and $Bi_2Te_3$, which are 3D TIs protected by time-reversal symmetry.[9-11] Recently, another class of topological insulators, called topological crystalline insulators (TCIs),[12] has been discovered in IV-VI semiconductors, such as SnTe, $Pb_{1-x}Sn_xSe$ and $Pb_{1-x}Sn_xTe$.[13-18] In TCIs, topological surface states are protected by the crystal symmetry inherent in these IV-VI semiconductors,[12, 13] and not by time reversal symmetry. Compared to topological insulators with time reversal symmetry, TCIs show several distinct materials differences. First, unlike $Bi_2Se_3$ and $Bi_2Te_3$ 3D TIs that possess a single surface state,[9] TCIs possess multiple surface states.[19] In SnTe for example, four topological surface states exist on each of the {100}, {110}, and {111} surfaces.[19] Moreover, the band structure of the surface states depends on the crystal symmetry of that particular surface.[17, 19] Second, because the surface states are protected by crystal symmetry and not by time reversal symmetry, they remain protected (gapless) in the presence of magnetic fields. Third, these IV-VI semiconductors can exhibit a wide range of electronic properties such as magnetism, superconductivity, and ferromagnetism, which are tunable by doping, strain, and alloying.[20-23] For example, In doping of SnTe leads to superconductivity, presenting a candidate topological superconductor.[24, 25] Thus, TCIs present opportunities to interface topological surface states with other types of electronic order. The distinct properties of TCIs thus make them an ideal platform for studying fundamental topological quantum phenomena that involve multiple surface states, the presence of magnetic fields, or interactions with ferromagnetism or superconductivity.

As different crystal surfaces contain their own topological surface states in TCIs, it is critical that the morphology of TCIs be controlled precisely to obtain particular topological surface states. Nanostructures are ideal to achieve this goal, as their morphology can be easily controlled by growth conditions and methods.[26, 27] In addition, nanostructures show clear facets with high crystalline quality and large surface-to-volume ratios, enhancing surface effects. Using a vapor-liquid-solid (VLS) growth technique with an Au catalyst, a few reports surveyed different morphologies of SnTe nanostructures, such as nanocrystals terminated with {100} surfaces only, or {100} and {110} surfaces, along with smooth or zigzag nanowires.[1-4] Here, we further probe the growth space and grow large SnTe nanoplates where the lateral dimension is in the micron range and the thickness is ~ 100 nm. SnTe nanoplates have not been demonstrated previously.

The advantage of nanoplates over nanowires is the large area of the top and bottom surfaces relative to the side surfaces. In this morphology, surface states of a particular crystal surface can be selectively enhanced. We find that the top and bottom surfaces of SnTe nanoplates are either {100} or {111}, depending on the growth substrate temperature. Thus, we can maximize surface states existing on {100} or {111} surfaces.

SnTe nanostructures were grown using a horizontal tube furnace. SnTe source powder (Alfa Aesar, 99.999% purity) was placed at the center of the tube furnace and heated to growth temperature (600 °C), while growth substrates were placed downstream in lower temperature zones (300 – 450 °C). $SiO_x$/Si substrates with or without a thin Au film were used for both VLS and vapor-solid (VS) growth modes. Figure 1(a) shows the schematic. During growth, flow of ultrapure Ar gas was maintained to transport the source vapor to the growth substrates. The SnTe source temperature, $SiO_x$/Si substrate temperature, base pressure, Ar gas flow rate, and growth time were varied to grow SnTe nanostructures with different morphologies. Depending on the substrate temperature and growth time, large area nanoplates, nanoribbons, or nanowires would dominate the growth products. Details of the growth conditions are provided in the Experimental section.

SnTe has a rock salt crystal structure, with each element having a face-centered cubic crystal structure, as shown in Fig. 1(b). Figure 1(c-k) shows different morphologies of SnTe nanostructures we obtain by controlling the growth substrate temperature. We divide SnTe nanostructures into four groups: nanoplates (Fig. 1(d), (e), (j)), nanoribbons (Fig. 1(f), (k)), nanowires (Fig. 1(g), (h), (i)), and nanoblocks (Fig. 1(c)). Au particles are observed at the end of nanoplates, nanoribbons, and nanowires, indicating VLS-promoted growth. However, the widths of the nanoplates and ribbons are much wider than the size of the Au particles, suggesting additional side growth via VS growth. Only the cross-sections of narrow nanowires (Fig. 1(g)) are similar in size to the Au catalyst. Without Au catalyst, nanoblocks were grown, which reflects the underlying rock salt crystal symmetry. All surfaces of nanoblocks are {100} planes. Optical images of the growth substrates and individual nanoplates with different facet angles are shown in Figure S1.

For nanoplates and nanoribbons, we observe that the top and bottom surfaces are either {100} or {111}. For plates or ribbons with {100} as the top surface (Fig. 1(d-f)), denoted as (100) nanoplates and ribbons from here on, the side facet angles are mostly 45, 90, or 135 degrees, indicating that side surfaces are {100} and {111}. For plates or ribbons with {111} as the top surface (Fig. 1(j), (k)), denoted as (111) nanoplates and ribbons, the side facet angles are 30, 60, or 120 degrees, indicating that the side surfaces are {100} and {111}. The crystal orientation of the top and bottom surfaces of nanoplates and ribbons is confirmed by transmission electron microscopy (TEM), which is discussed below. We note that (110) nanoplates or nanowires are not observed in our growth products, in agreement with the previous report that the surface energy of {110} planes is higher than those of {100} or {111}.[4] For nanowires, they grow along the <100> direction with {100} side surfaces. Two types of nanowires are observed: one shows a constant cross-section along the wire (Fig. 1(g, h)) and the other a decreasing cross-section along the wire (Fig. 1(i)). The nanowire shown in Fig. 1(g) is very narrow, with a diameter less than 100nm.

The morphology of SnTe nanostructures depends on the growth substrate temperature. The substrate temperature of (100) nanoplates and nanoribbons, and <100> nanowires is roughly 350-450°C, while that of (111) nanoplates and nanoribbons is around 300°C. Thus, {111}- terminated SnTe nanostructures grow at lower temperatures than {100}-terminated SnTe nanostructures, consistent with the reported results for {111} nanocrystals. [4]

We note that different topological surface states are present on {100} and {111} surfaces of SnTe. On {100} surfaces, four Dirac-dispersive surface states exist in the 1$^{st}$ Brillouin zone where the Dirac point is slightly displaced from the X point.[14, 19] Two surface states, each located on either side of the X point, merge together at energies away from the Dirac point. Four surface states also exist on {111} surfaces: one centered at the Γ point and three at each M point.[17, 19] Because of the different crystal symmetry in (100) and (111) surfaces, the band structure of these surface states is different. Therefore, (100)- and (111)-nanoplates with large lateral dimensions provide opportunities to selectively study transport properties of particular surface states.

We characterize the atomic structure and the Sn:Te elemental ratio of our SnTe nanostructures using TEM and scanning TEM (STEM) energy dispersive X-ray spectroscopy (EDX). Figure 2 shows TEM results for SnTe nanoplates and nanoribbons. For (100) nanoplates and ribbons, selected area diffraction (SAED) patterns clearly show cubic symmetry of the (100) projection, as shown in the insets of Fig. 2(a) and 2(d). High resolution TEM images show square lattice fringes with a spacing of ~ 0.32 nm, corresponding to (200) spacing (Fig. 2(c) and 2(e)). The edge of the (100) nanoribbon, shown in Fig. 2(d), is partially terminated with a (110) plane, as indicated by the lattice spacing of ~ 0.23 nm (Fig. 2(e)). STEM-EDX mapping was carried out on the SnTe nanoribbon to show even distributions of Sn and Te, as expected (Fig. 2(f)). Au catalyst is found at the end of the ribbon. A (111) SnTe nanoplate is shown in Fig. 2(g). The diffraction pattern (Fig. 2(h)) shows the hexagonal symmetry expected in the (111) projection, and the lattice spacing of ~ 0.23 nm (Fig. 2(i)) represents (220) spacing. Au, Sn, and Te elemental maps are shown in Fig. 2(j). Supplementary Figure S2 shows EDX spectra of SnTe nanostructures shown in Fig. 2, confirming that the Sn:Te ratio is 1:1.

We measure four point resistance values of SnTe nanoplates, ribbons, and wires to survey if the electrical properties depend on the morphology and the crystallographic orientation of SnTe. Figure 3(a) and 3(b) show the linear I-V curve of a SnTe nanowire and (100) nanoplate, respectively. Irrespective of morphological differences, their resistance values are low, < 100 Ω, which suggest high carrier concentrations. The cross section of the nanowire device shown in Fig. 3(a) actually decreases along the length of the wire. This is reflected in the increase in the four-point resistance values and shown clearly in the inset images. From four point resistance measurements, we obtain room temperature resistivity values of different SnTe nanostructures. This is shown in Fig. 3(c). (Table S1 reports resistance, thickness, and resistivity values of all the devices, measured at room temperature.) Resistivity values fall mostly between 100 and 1000 μΩ-cm for all SnTe nanostructures regardless of the nanostructure shape or crystallographic orientation.

SnTe is known to be a heavily doped p-type semiconductor due to Sn vacancies.[28, 29] We measure the carrier concentration of (111)- and (100)- nanoplates by fabricating Hall bar devices with thermally evaporated Cr/Au as electrodes. We first track the longitudinal resistance, $R_{xx}$, down to 2 K. Hall measurements ($R_{yx}$) are carried out at 2 K. Figure 4(a) shows temperature-dependent $R_{xx}$ of a (111) nanoplate. We observe a discernible kink at ~ 47 K. This anomaly in resistance is due to a structural phase transition from a high temperature rock salt to a low temperature rhombohedral structure[30, 31] that is related to a ferroelectric phase transition.[32] The 2D hole carrier density from the Hall measurement (Fig. 4(b)) is $2.4 \times 10^{15}$ cm$^{-2}$. Given that the thickness of the plate is 107 nm, its bulk carrier density is $2.9 \times 10^{20}$ cm$^{-3}$. The transition temperature of ~ 47 K is also observed in SnTe bulk samples with a carrier density of ~$10^{20}$ cm$^{-3}$.[31] We note that unlike the SnTe thin film, which showed non-linear Hall resistance suggesting multiple conducting channels,[33] the Hall resistance of the (111) nanoplate measured here is linear. This indicates that bulk transport is dominant in our SnTe nanoplates, with a measured Hall mobility of 670 cm$^2$/Vs. Figure 4(c) shows $R_{xx}$ of a (100) nanoplate down to 2 K. There are no kinks in the resistance, which decreases linearly with temperature (bottom inset of Fig. 4(c)), showing metallic behavior similar to topological insulators with high bulk carrier density.[34] From the Hall data, we find the bulk hole carrier density to be 9.8x $10^{21}$ cm$^{-3}$, one order of magnitude higher than that of the (111) nanoplate shown in Fig. 4(a,b). It is known that the phase transition temperature decreases with increasing carrier density.[31] Therefore, we conclude that the phase transition temperature must be lower than 2 K, explaining the absence of a kink in $R_{xx}$. The higher carrier density for the (100) nanoplate compared to the (111) nanoplate may be related to the higher substrate temperature for the (100) nanoplate, which would induce more Sn vacancies. The Hall slope for this device is nonlinear, as indicated by the departure of the Hall resistance from the linear fit based on the low magnetic fields up to 3 T. (Fig. 4(d)). This indicates the existence of multiple conduction channels with different mobilities. We did not try a two-channel model (one bulk channel and the other surface channel) to fit the nonlinear Hall curve as it is unclear how many channels are present here. The average Hall mobility for this nanoplate is 10.5 cm$^2$/Vs, roughly estimated from the low field linear fit. This mobility is much lower than that of the first nanoplate (Fig. 4(a,b)), likely due to the much higher carrier density. Another nanoplate with even higher carrier density is shown in Fig. 4(e,f). $R_{xx}$ shows a slight upturn at low temperature below 10 K (the bottom inset of Fig. 4(e)). The carrier density, deduced from the linear fit of the Hall curve up to 3 T (Fig. 4(f)), is 3.9 x $10^{22}$ cm$^{-3}$. At such a high carrier density, the screening length is small, as it is proportional to $n^{-1/6}$ where n denotes the carrier density.[35] Thus, electron-electron interactions play a role, which can explain the slight increase of $R_{xx}$ at low temperature. The linear dependence of the longitudinal conductance, $\sigma_{xx}$, on ln(T), shown in the top inset of Fig. 4(e), further confirms electron-electron interactions.[36,34] Differing from topological insulators with strong spin-orbit coupling as well as weak antilocalization at low magnetic fields,[34, 36] the resistance of this nanoplate keeps the same increasing slope with ln(T) at different magnetic fields. This is demonstrated by the nearly overlapping points of $\Delta\sigma_{xx}$ at different magnetic fields, which are displayed in the inset of Fig. 4(f). Again, no kinks are observed in $R_{xx}$ down to 1.8 K, which is attributed to the high carrier density.

The structural phase transition observed at ~ 47 K for the (111) nanoplate means that the mirror symmetry in the (100) plane is lost, while the mirror symmetry in the (111) plane is intact. Thus, topological surface states present on {100} surfaces of such nanoplates may be gapped while surface states on {111} surfaces are still protected. However, the structural phase transition is small, and thus the bandgap opening in the surface state is expected to be minimal. The structural phase transition presents the coexistence of ferroelectricity with topological surface states, which may be interesting for future study.

The carrier concentrations of the nanoplates studied here are large, consistent with other intrinsic SnTe bulk crystals and thin films.[28, 33, 37] The transport values (carrier density and mobility) are similar to values of intrinsic $Bi_2Te_3$ nanoplates,[26] which are heavily doped n-type semiconductors due to anti-site defects. To study the topological surface states present on {100} and {111} surfaces in (100) and (111) SnTe nanoplates respectively, the bulk carrier concentration must first be reduced. In order to decrease the bulk carrier concentration, we may grow the SnTe in a Sn-rich vapor environment, or dope the SnTe with Bi, Pb, or Sb donors.[18, 29, 34, 36-38]

As a new class of topological insulators, TCIs exhibit multiple Dirac-dispersive surface states that are protected by crystal symmetry, not time-reversal symmetry. The presence of multiple surface states, crystal surface dependent band structures, and protection against magnetic fields offer new opportunities to study topological quantum phenomena that are inaccessible in 3D topological insulators with time-reversal symmetry. Here, we have presented different SnTe nanostructure morphologies grown by VLS and VS growth modes. In particular, we have grown SnTe nanoplates with {100} or {111} as the top and bottom surfaces. At present, the SnTe nanoplates show high bulk carrier concentrations due to Sn vacancies. A kink in resistance at ~ 47 K indicates a structural transition from a rock salt to a rhombohedral structure. In addition, electron-electron interactions occur in nanoplates with high carrier density. Growth in a Sn-rich environment or donor doping may be necessary to reveal the surface states in these SnTe nanoplates. Notably, the large {100} or {111} surface area of SnTe nanoplates allows selective study of surface states present on these surfaces.

**Experimental Methods**

*SnTe Nanostructure Growth.* SnTe nanostructures were grown using a horizontal tube furnace with a 1-inch diameter quartz tube. The tube was flushed with ultrapure Ar gas several times before growth to remove residual oxygen and moisture. The base pressure of the tube during growth was maintained at 2 – 5 Torr. The growth temperature at the center of the tube furnace was 600 °C, and the substrate temperature ranged between 300 °C and 450 °C. The ramping time to reach the growth temperature was 30 minutes and the temperature was maintained for 10 minutes, followed by a natural cool down. During the entire growth, ultrapure Ar was delivered at a flow rate of 40 – 50 standard cubic centimeter per minute. (100) nanoplates and ribbons were found on substrates located at 350 – 450 °C. (111) nanoplates and ribbons were found on substrates located at ~ 300 °C. For nanowires, the substrate temperature was roughly 350 °C. Small {100} nanoplates and ribbons were also found on the lowest temperature substrate (300 °C).

*SnTe Nanodevice Transport.* Nanodevices were fabricated on SnTe nanoplates and nanowires grown on 300 nm $SiO_2$/Si. E-beam lithography was used to pattern electrodes and 10/150nm Cr/Au was thermally evaporated for the electrodes. For low temperature magnetotransport measurements of SnTe nanodevices, a Quantum Design PPMS Dynacool was used. We measured the longitudinal resistance down to 1.8 K. Hall measurements were carried out in magnetic fields up to 9 T at 1.8 K.

FIGURES CAPTIONS

**Figure 1.** Synthesis of SnTe nanostructures. (a) Growth schematic. SiO$_2$/Si substrates are placed downstream in the tube furnace. The center of the tube furnace is heated to 600 °C. Substrate temperatures range between 300 °C and 450 °C. The morphology of SnTe nanostructures is sensitive to the substrate temperature. (b) SnTe unit cell. (c) {100} cubic crystals, grown without Au catalyst. (d, e) (100) nanoplates. (f) (100) nanoribbon. (g, h, i) Nanowires with <100> growth direction. (j) (111) nanoplate and (k) (111) nanoribbon. At the highest substrate temperatures (350 – 450 °C), dominant growth products are (100) nanoplates and nanoribbons. Nanowires are found at ~ 350 °C substrate temperature. At lowest substrate temperatures (~300 °C), (111) nanoplates and ribbons are found. Thin (100) nanoplates and ribbons are also found. Scale bars are 2μm (solid, white lines) and 200 nm (dashed, yellow lines). Facets are marked on the structures.

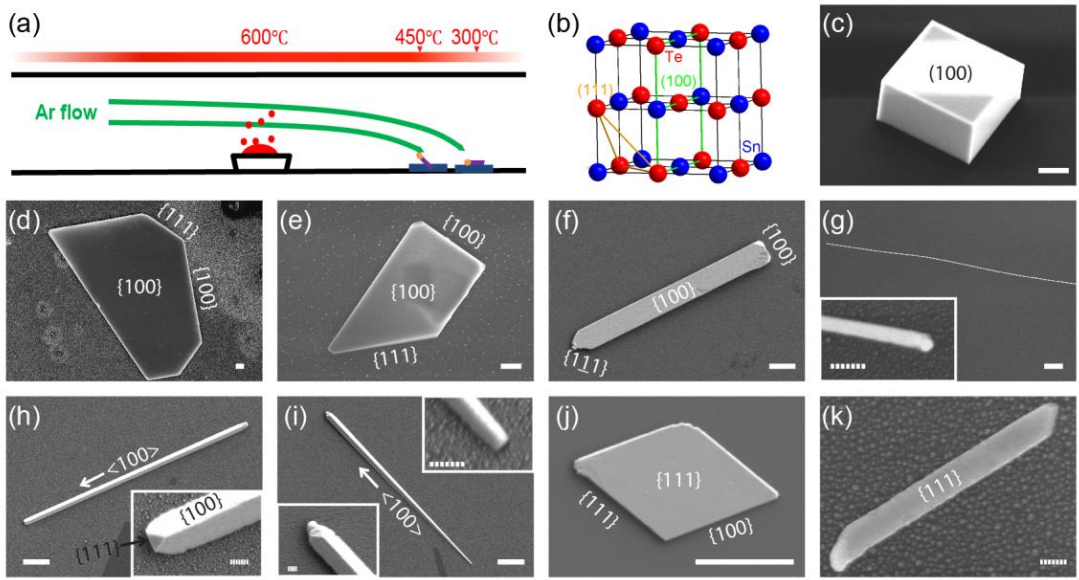

**Figure 2.** Atomic structure characterization of (100) and (111) SnTe nanoplates and ribbons. (a) (100) nanoplate. The diffraction pattern (bottom inset) shows a square symmetry, indicating that the top and bottom surfaces are (100). (b, c) show lattice fringes with a lattice spacing of ~ 3.18 Å, which corresponds to (200) spacing. (b) is a close-up view of the red boxed region in (a) and (c) is a close-up view of the red boxed region in (b). (d) (100) nanoribbon with (100) diffraction pattern (inset). (e) shows lattice spacing of 2.32 Å, corresponding to (220) spacing. (f) Elemental maps of Sn, Te, and Au, where Au is located at the tip of the nanoribbon, indicating VLS-assisted growth. Scale bars = 60nm. (g) (111) nanoplate. The hexagonal symmetry in the diffraction pattern (h) shows that the top and bottom surfaces are (111). (i) The lattice spacing of ~ 2.3 Å corresponds to (220) spacing. (j) HAADF and elemental maps of the (111) nanoplate. Scale bars = 300nm. Scale bars in the three diffraction patterns are 10.0/Gm.

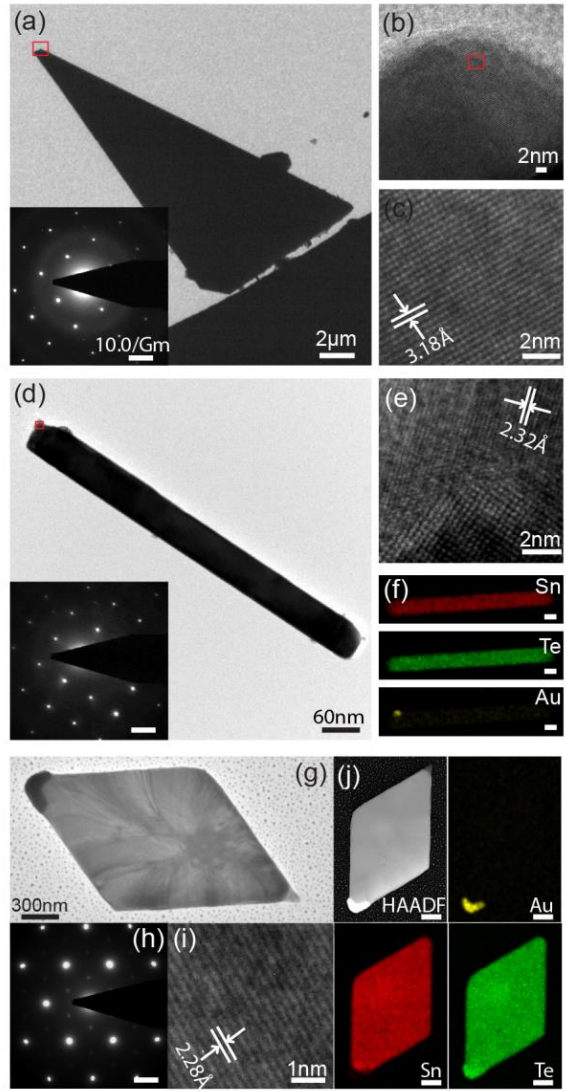

**Figure 3.** Resistivity range of various SnTe nanostructures. (a) I-V curves of a typical nanowire with decreasing diameter. Top inset: SEM image of the corresponding device. Bottom inset: Two segments (different colors in the top inset) that were measured. Scale bars=500nm. The I-V curve in blue is from the thinner segment of the nanowire. As expected, the resistance is ~ twice as high for the thinner segment. (b) Linear I-V curve of a typical (100) nanoplate. The inset shows the corresponding device. (c) Resistivities of various SnTe nanostructures, including (100)

and (111) nanoplates and nanoribbons, and <100> nanowires. Most nanostructures have resistivities between 100 and 1000 μΩ·cm, suggesting high carrier concentrations.

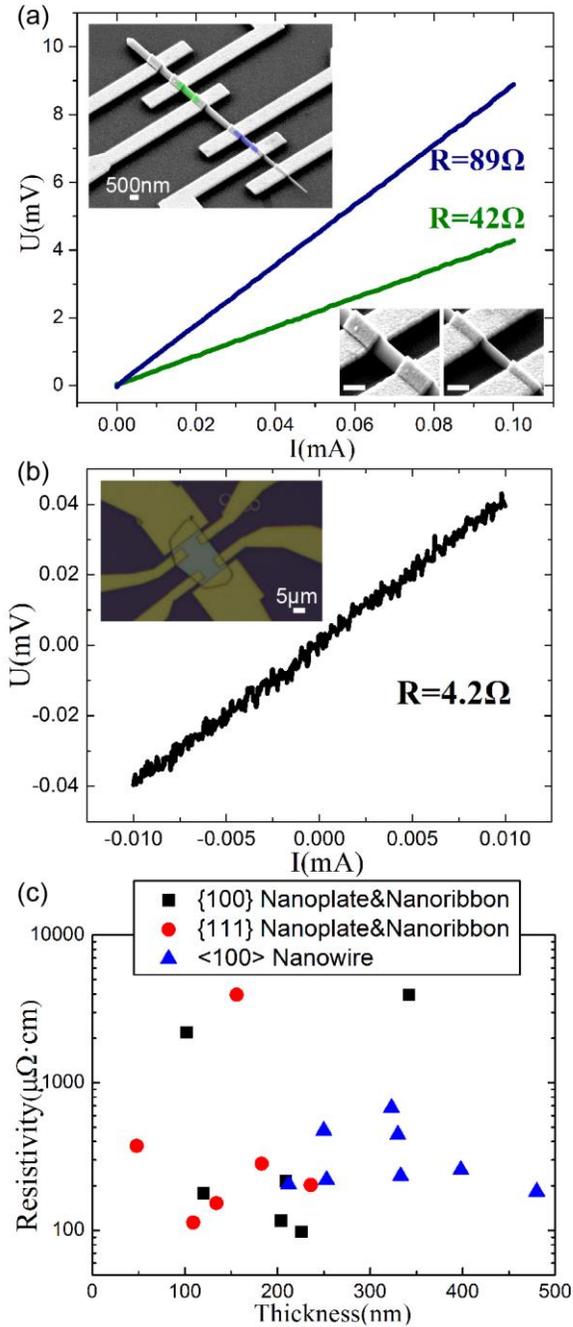

**Figure 4**. Low temperature magnetotransport of (100) and (111) nanoplates of different carrier densities. (a) $R_{xx}$ vs. T and (b) $R_{yx}$ vs. B of a (111) nanoplate device (inset in (a)). A kink is observed in $R_{xx}$ at ~ 47 K (red arrow), which indicates a structural phase transition. The Hall slope fits well to a linear curve. (c) $R_{xx}$ vs. T and (d) $R_{yx}$ vs. B of a (100) nanoplate (inset in (c)). $R_{xx}$ decreases monotonically with temperature, not showing any kinks down to 1.8 K. This is due to a higher carrier density, measured from the Hall curve in (d). The Hall slope is linear at low magnetic fields, but non-linear at high fields, indicating multiple conduction channels. (e) $R_{xx}$ vs. T of another nanoplate with a much higher carrier density. At low temperatures (< 10 K), the resistance has a slight upturn (bottom inset of (e)), and its conductance vs ln(T) curve is nearly linear (top inset of (e)), suggesting electron-electron interactions. (f) $R_{yx}$ vs. B of the device in (e), showing a non-linear Hall resistance and a very high carrier density. The inset shows that the longitudinal conductance, $\Delta\sigma_{xx}=\Delta\sigma_{xx}(T)-\Delta\sigma_{xx}(1.8K)$, measured at different magnetic fields falls on the same curve, scaling linearly with ln(T) with the same slope. This suggests electron-electron interactions due to the high carrier density.

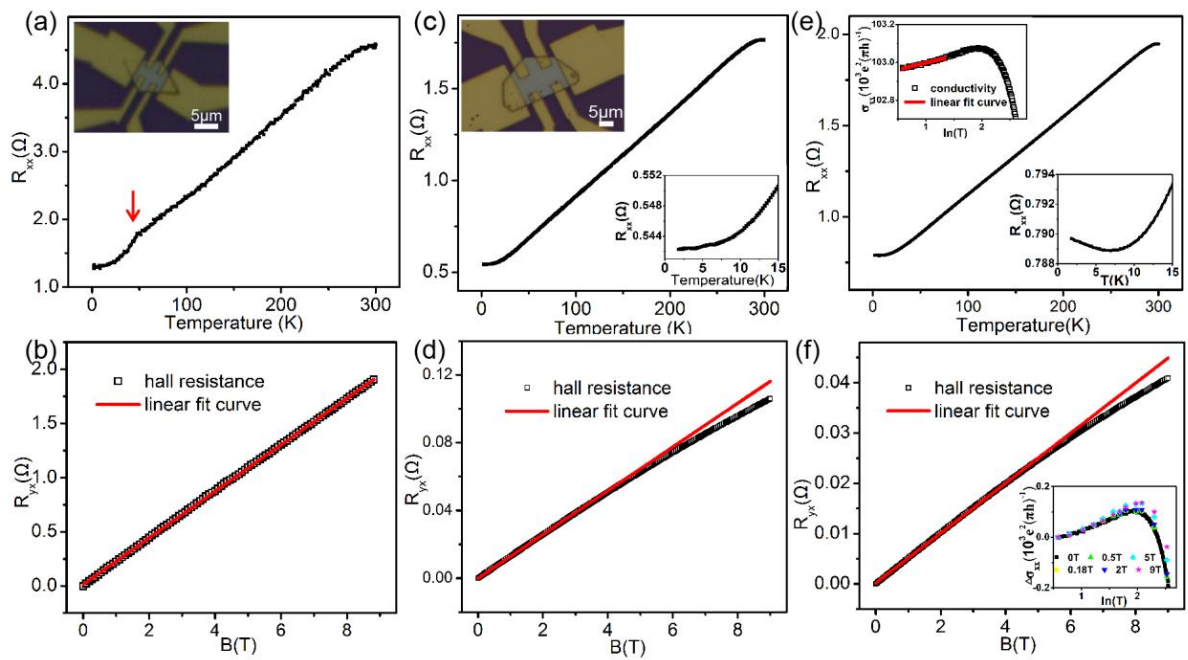

## ASSOCIATED CONTENT

**Supporting Information**. Optical images, energy dispersive X-ray spectra, and resistivity values of SnTe nanostructures are available in supporting information. This material is available free of charge via the Internet at http://pubs.acs.org.

## AUTHOR INFORMATION

**Corresponding Author**

* Judy J. Cha, E-mail: judy.cha@yale.edu

**Notes**

The authors declare no competing financial interest.

**Author Contributions**

The manuscript was written through contributions of all authors. All authors have given approval to the final version of the manuscript.

ACKNOWLEDGMENT

Microscopy facilities used in this work were supported by the Yale Institute for Nanoscience and Quantum Engineering and National Science Foundation MRSEC DMR 1119826. The Quantum Design PPMS Dynacool used in this work is supported by National Science Foundation MRSEC DMR 1119826 (CRISP).